\documentstyle[12pt]{article}

\def\d{{\partial}}
\newcommand{\be}{\begin{equation}}
\newcommand{\ee}{\end{equation}}
\newcommand{\ba}{\begin{eqnarray}}
\newcommand{\ea}{\end{eqnarray}}
\newcommand{\no}{\nonumber}
\newcommand{\al}{\alpha}
\newcommand{\bt}{\beta}
\newcommand{\gm}{\gamma}
\newcommand{\sg}{\sigma}
\newcommand{\la}{\langle}
\newcommand{\ra}{\rangle}
\newcommand{\Op}{{\cal O}}

\begin{document}

\thispagestyle{empty}

\begin{flushright}
UT-769, YITP-97-8 \\
LBNL-40098 \\
March, 1997 \\
hepth/9703086\\
\end{flushright}

\bigskip

\begin{center}
{\Large \bf Quantum Cohomology and Virasoro Algebra}
\end{center}

\bigskip

\bigskip

\begin{center}

Tohru Eguchi 

\medskip

{\it Department of Physics, University of Tokyo, 

\medskip

Tokyo 113, Japan}

\bigskip

\bigskip

Kentaro Hori

\medskip

{\it Department of Physics, University of California, Berkeley, 

\medskip

CA 94720-7300 U.S.A.}

\bigskip

\bigskip

and

\bigskip

\bigskip

Chuan-Sheng Xiong

\medskip

{\it Yukawa Institute for Theoretical Physics, Kyoto University 

\medskip

Kyoto 606, Japan}
\end{center}

\bigskip

\bigskip

\begin{abstract}
We propose that the Virasoro algebra controls quantum cohomologies of
general Fano manifolds $M$ ($c_1(M)>0$) and determines their partition
functions at all genera. We construct Virasoro operators in the case of
complex projective spaces and show that they reproduce the results
of Kontsevich-Manin, Getzler etc. on the genus-0,1 instanton numbers. 
We also construct Virasoro operators for a wider class of Fano varieties.
The central charge of the algebra is equal to $\chi(M)$, the Euler 
characteristic of 
the manifold $M$.
\end{abstract}
    
\newpage
\pagenumbering{arabic}

As is well known, the quantum cohomology of a (symplectic) manifold $M$ 
is described
by the topological $\sg$-model with $M$ being its target space \cite{Wa}.
The partition function of the topological $\sg$-model is given by the 
sum over 
holomorphic maps from a Riemann surface $\Sigma_g$ to $M$. When the 
degree 
$d$ of the map is zero (constant map), correlation functions of the 
$\sg$-model 
reproduces the classical intersection numbers among homology cycles of 
$M$.
When the degree is non-zero, however, the theory describes the quantum
modification of the classical geometry due to the presence of the 
world-sheet 
instantons. In the following we consider the case where the topological
$\sg$-model is coupled to two-dimensional gravity so that we consider 
Riemann
surfaces of an arbitrary genus and incorporate the gravitational 
descendants.

Recently there have been extensive studies of topological $\sg$ models 
coupled to two-dimensional gravity (Gromov-Witten invariants) 
\cite{KM,Itz,DI,
Du,Kon,Curves,RT,McS}.
In ref.\cite{EHX} superpotentials for a large class of Fano manifolds $M$ 
(complex projective spaces, Grassmannians, rational surfaces etc.) have 
been constructed so that they reproduce Gromov-Witten invariants by 
means of residue integrals
(see \cite{Givena} for a related analysis using oscillatory 
integrals).
Existence of the superpotentials indicates the mirror 
phenomenon for Fano varieties analogous to the one in the case 
of Calabi-Yau manifolds. These results, however, have so far been 
limited to
the genus-0 case and no general principle is known to control quantum 
cohomology
at higher genera (recursion relation for genus-1 instanton numbers has
recently been obtained by Getzler \cite{Getz}). 

In this article we would like to propose a new powerful algebraic 
machinery 
for organizing the quantum cohomology theory at all genus: we propose that 
the Virasoro algebra controls the quantum cohomology and determines their 
partition functions at an arbitrary genus. We first discuss the 
Virasoro generators in the case of $CP^N$ manifolds and show that the
Virasoro conditions eq.(1) correctly reproduce the results on instanton 
numbers of projective spaces obtained by
\cite{KM,Itz,DI,JS,Getz} using the associativity of quantum cohomology ring. 
We then present the
general form of the Virasoro operators for a wider class of Fano varieties.
It turns out that the central charge of the Virasoro algebra equals the
Euler characteristic of the manifold $M$.

Our conditions take the form
\be
L_nZ=0, \hskip20mm n=-1,0,1,2,\cdots
\ee
where $Z$ is the partition function related to the free energy of the 
theory as 
\be
Z=\exp\big(\sum_{g=0}\lambda^{2g-2}F_g\big)
\ee
$\lambda$ is the genus expansion parameter. Operators $L_n$ form the
Virasoro algebra $[L_n,L_m]=(n-m)L_{n+m}$.  

Cohomology classes of the complex projective space $M=CP^N$ is given by
$\{1,\omega,\omega^2,\cdots,\omega^N\}$ where $\omega$ is the 
K\"{a}hler class. Corresponding fields are denoted as
$\{{\cal O}_{\alpha}, \alpha=0,1,2,\cdots,N\}$ and their coupling 
parameters are 
given by $\{t^{\al}\}$. Gravitational descendants of $\cal{O}_{\al}$
are denoted as $\sg_n(\cal{O}_{\al})$,$\hskip1mm n=1,2,\cdots$ 
and their parameters by $\{t_{n}^{\al}\}$.
Virasoro operators for $CP^N$ are then defined by
\ba
&&\hskip-8mmL_{-1}=\sum_{\alpha=0}^N \sum_{m=1}^\infty 
mt_m^\alpha\d_{m-1,\alpha}
+\frac{1}{2\lambda^2}\sum_{\alpha=0}^N t^\alpha t_\alpha, \\
&&\hskip-8mmL_0=\sum_{\alpha=0}^N \sum_{m=0}^\infty 
(b_\alpha+m)t_m^\alpha\d_{m,\alpha}
+(N+1)\sum_{\alpha=0}^{N-1} \sum_{m=0}^\infty mt_m^\alpha\d_{m-1,\alpha+1}
+\frac{1}{2\lambda^2}\sum_{\alpha=0}^{N-1} (N+1)t^\alpha t_{\alpha+1}\no\\
&&-\frac{1}{48}(N-1)(N+1)(N+3), \\
&&\hskip-8mm L_n=\sum_{m=0}^\infty\sum_{\alpha=0}^N 
\sum_{j=0}^{N-\alpha}
(N+1)^jC_\alpha^{(j)}(m,n) t_m^\alpha\d_{m+n-j,\alpha+j} \hskip20mm n\ge 1\\
&&\hskip-5mm+\frac{\lambda^2}{2}\sum_{\alpha=0}^N
\sum_{j=0}^{N-\alpha}\sum_{m=0}^{n-j-1}
(N+1)^jD_\alpha^{(j)}(m,n) \d^\alpha_m\d_{n-m-j-1,\alpha+j}
+\frac{1}{2\lambda^2}\sum_{\alpha=0}^{N-n-1} 
(N+1)^{n+1}t^\alpha t_{\alpha+n+1},\no
\ea
where
\ba
C_\alpha^{(j)}(m,n)\hskip-2mm&\equiv&\hskip-2mm 
\frac{(b_\alpha+m)(b_\alpha+m+1)
\cdots(b_\alpha+m+n)}{(m+1)(m+2)\cdots(m+n-j)}\no\\
&\times&
\sum_{m\le \ell_1<\ell_2<\cdots<\ell_j\le m+n}
\bigg(\prod_{i=1}^j \frac{1}{b_\alpha+\ell_i}\bigg),
\ea
and
\ba
D_\alpha^{(j)}(m,n)\hskip-2mm&\equiv&\hskip-2mm 
\frac{b^\alpha(b^\alpha+1)\cdots(b^\alpha+m)
b_\alpha(b_\alpha+1)
\cdots(b_\alpha+n-m-1)}{m!(n-m-j-1)!}\no\\
&\times&\sum_{-m-1\le \ell_1<\ell_2<\cdots<\ell_j\le n-m-1}
\bigg(\prod_{i=1}^j \frac{1}{b_\alpha+\ell_i}\bigg).
\ea
Here $\d_{n,\al}=\d/\d t_{n}^{\al}$ and $b_{\al}$ is related to the 
degree $q_{\al}$ of the cohomology class 
$\omega_{\al}$ as
\be
b_\alpha\equiv q_\alpha-\frac{1}{2}(N-1),\qquad
\omega_{\al}\in H^{2q_{\al}}(M).
\ee
The indices are raised and
lowered by the metric
\be
\eta^{\alpha\beta}=\delta_{\al+\bt,N}
\ee
as usual. Thus $q^{\al}=N-q_{\al}, \hskip1mm b_{\al}+b^{\al}=1$.

Note that the factor $N+1$ which appears in the right-hand-side of 
eqs.(4),(5)  
is the magnitude of the first Chern class of $CP^N, c_1(CP^N)
=(N+1)\omega$. Thus the terms $j\not= 0$ represent the broken scale 
invariance of the $\sg$-model with the target manifold $M$ which is 
a Fano 
variety. If one considers a fictitious manifold with a vanishing first 
Chern class with dimension $k/k+2$, then it turns out that the
above Virasoro operators 
reduce to the well-known expressions in the theory of two-dimensional
gravity and KP hierarchy \cite{DVV}. Thus eqs.(3-5) may be regarded as 
the generalization
of Virasoro conditions of 2-dimensional gravity to the case of the topological
$\sg$-model.
We also note that when $N=1$ the above expressions reduce to those
already obtained in \cite{EHY}.

Virasoro algebra
\be
[L_n,L_m]=(n-m)L_{n+m}, \qquad n,m\ge-1,
\ee
can be verified by making use of the identities
\ba
&&\sum_{j=0}^i
C_\alpha^{(i-j)}(k,n)C^{(j)}_{\alpha+i-j}(k+n-i+j,m) \no \\
&&=(b_\alpha+k+n)C_\alpha^{(i)}(k,n+m)+(k+n+m-i+1)C_\alpha^{(i-1)}(k,n+m), \\
&&\sum_{j=0}^iD_\alpha^{(i-j)}(k,n)C^{(j)}_{\alpha+i-j}(n-k-i+j-1,m) \no \\
&&=(b_\alpha+n-k-1)D_\alpha^{(i)}(k,n+m)+(n+m-k-i)D_\alpha^{(i-1)}(k,n+m) 
\ea

We have made an extensive check of the Virasoro conditions eq.(1).
For the sake of illustration we consider the case of $CP^2$ and 
$L_1Z=0$ equation. We denote the three primary fields corresponding to 
the cohomology classes $1,\omega,\omega^2$ as $P,Q,R$ and denote their 
couplings as $t^P,t^Q,t^R$. Genus-0 free energy has an expansion
\ba
&&F_0={1 \over 2}t^P(t^Q)^2+{1 \over 2}(t^P)^2t^R+f(t^Q,t^R),  \\
&&f(t^Q,t^R)=\sum_{d=1}N_d^{(0)}{(t^R)^{3d-1} \over (3d-1)!}
\exp(d\hskip1mmt^Q)
\ea
where $N_d^{(0)}$ is the number of genus-0 instantons of degree $d$
(number of degree $d$ rational curves passing through $3d-1$ points).

In the small phase space with $t^{\al}_n=0$ for all $\al$ and 
$n\ge 1$ except for $t^P_1=-1$, $L_1Z=0$ equation reads as
\ba
&&-{3 \over 8}\la\sg_2(P)\ra_0+{15 \over 4}t^R\la\sg_1(R)\ra_0
+{3 \over 4}t^Q\la\sg_1(Q)\ra_0-{1\over4}t^P\la\sg_1(P)\ra_0
-6\la\sg_1(Q)\ra_0\no \\
&&+6t^Q\la R\ra_0 -9\la R\ra_0
-{3 \over 4}\la P\ra_0\la R\ra_0+{1 \over 8}\la Q\ra_0\la Q\ra_0
+{9 \over 2}(t^P)^2=0.
\ea
$\la \cdots \ra_0$ denotes the genus-0 correlation function.
We take the second derivative of (15) in $t^Q$  and set $t^P=t^Q=0$.
We obtain
\ba
&&-{3 \over 8}\la\sg_2(P)QQ\ra_0+{15 \over 4}t^R\la\sg_1(R)QQ\ra_0
+{3 \over 2}\la\sg_1(Q)Q\ra_0-6\la\sg_1(Q)QQ\ra_0 \\
&&+12\la RQ\ra_0-9\la RQQ\ra_0
-{3 \over 4}\la PQQ\ra_0\la R\ra_0-{3 \over 2}\la PQ\ra_0\la RQ\ra_0
-{3 \over 4}\la P\ra_0\la QQR\ra_0 \no \\
&&+{1 \over 4}\la QQ\ra_0\la QQ\ra_0
+{1 \over 4}\la QQQ\ra_0\la Q\ra_0=0. \no
\ea
In order to eliminate descendant fields in (16) we use the topological
recursion relation (TRR) at genus-0 \cite{Wb},
\be
\la\sg_n(\Op_{\al})XY\ra_0=n\la\sg_{n-1}(\Op_{\al})\Op_{\bt}\ra_0
\la\Op^{\bt}XY\ra_0
\ee
($X,Y$ are arbitrary fields) and also the "flow equations"
\ba
&&\la\sg_1(P)P\ra_0=uw+{1 \over 2}v^2, \hskip10mm \la\sg_1(P)Q\ra_0
=vw-f_v+uf_{uv}+vf_{vv} \\
&&\la\sg_1(P)R\ra_0={1 \over 2}w^2+uf_{uu}+vf_{uv}-f_u,
\hskip3mm \la\sg_1(Q)Q\ra_0=vf_{uv}+{1 \over 2}(w+f_{vv})^2 \no
\ea
where $u,v,w$ are defined as
\be
u=\la PP\ra_0=t^R, \hskip2mm v=\la PQ\ra_0=t^Q, \hskip2mm w=\la PR\ra_0=t^P
\ee
and $f_u=\d f/\d u$ etc. Flow equations are derived using TRR.
Then eq.(16) is rewritten as
\ba
&&3t^R\la RR\ra_0+6\la QR\ra_0-9\la QQR\ra_0+\la Q\ra_0\la QQQ\ra_0
+3t^R\la QR\ra_0 \la QQQ\ra_0 \no\\
&&-6\la QQ\ra_0\la QQQ\ra_0+(\la QQ\ra_0)^2=0.
\ea
By using the expansion of the free energy (14) we can convert (20) 
into a relation for the instanton numbers. After some algebra we find
\be
N^{(0)}_d=(3d-4)!\sum{N_{\ell}^{(0)}N_{k}^{(0)} \over (3\ell-1)!(3k-1)!}
k^2\ell[3k\ell+\ell-2k]
\ee
which is the well-known result of Kontsevich and Manin \cite{KM}.

The above procedure may be made more systematic as follows:
it is possible to show that in the small phase space
and at genus=0, the $L_1Z=0$ condition may be 
rewritten as
\be
A^\al A^\beta
\Bigl[\la \Op_\al \Op_\mu\Op_\gamma\ra_0\,\,
\la \Op^\gamma\Op_\beta \Op_\nu\ra_0\,
-\la \Op_\al \Op_\beta\Op_\gamma\ra_0\,\,
\la \Op^\gamma\Op_\mu\Op_\nu\ra_0\Bigl]=0,
\ee
where
\be
A^{\al}\equiv (q_{\al}-1)t^{\al}-({\cal C})_0^{\hskip2mm \al}.
\ee
${\cal C}$ is the matrix representation of the first Chern class
\be
({\cal C})_{\al\bt}\equiv \int_{M} c_1\wedge\omega_{\al}\wedge\omega_{\bt}.
\ee
Repeated indices ($\al,\bt,\gm$) are summed in (22). In this form the $L_1$ 
equation has the structure of the genus-0 associativity equation and it 
reproduces the results based on the associativity of quantum cohomology
ring \cite{KM,Itz,DI,JS}. 
  
We may also consider the genus-1 instanton numbers and check against the 
recent results of ref. \cite{Getz,CH}. Genus-1 free energy of $CP^2$ 
has an expansion
\be
F_1={-t^Q \over 8}+\sum_{d=1} N_d^{(1)}{(t^R)^{3d} \over (3d)!}
\exp(d \hskip1mm t^Q).
\ee
We again consider the equation $L_1Z=0$ for simplicity. 
This time we use the TRR 
at genus-1 \cite{Wb}
\be
\la \sigma_n({\cal O}_{\alpha})\ra_1={n \over 24}
\la \sigma_{n-1}({\cal O}_{\alpha}){\cal O}_{\beta}{\cal O}^{\beta}\ra_0
+n\la \sigma_{n-1}({\cal O}_{\alpha}){\cal O}_{\beta}\ra _0
\la {\cal O}^{\beta}\ra_1
\ee
and the flow equation eq.(18). After some algebra we find
\ba
&&\la Q\ra_0\la Q\ra_1+3t^R\la QR\ra_0\la Q\ra_1+{1 \over 8}t^R\la QQR\ra_0
-{1 \over 4}\la QQQ\ra_0 \no \\
&&-6\la QQ\ra_0\la Q\ra_1+{1 \over 8}\la QQ\ra_0
-9\la R\ra_1=0
\ea
where $\la\cdots\ra_1$ denotes genus-1 correlators.
Using the free-energy expansion (25) we find 
\be
N_n^{(1)}=N_n^{(0)}{1 \over 72}n(n-1)(n-2)+\sum_{k+\ell=n}(3n-1)!
{N_{k}^{(0)} \over (3k-1)!}{N_{\ell}^{(1)} \over (3\ell)!}{\ell \over 9}
(3k^2-2k).
\ee
The above equation has a form somewhat different from the one of \cite{Getz}, 
however, predicts the same instanton numbers. We can also check that the 
genus-1 instanton numbers of $CP^3$ are correctly reproduced by Virasoro 
conditions.

Let us next describe the derivation of the Virasoro conditions eqs.(1)--(5).
$L_{-1}$ is the well-known string equation which has a universal form
for all manifolds $M$ \cite{DW}. $L_0$ operators has been obtained in \cite{Ho}
by using the
intersection theory on the moduli space of Riemann surfaces. Its general
form is given by
\ba
L_0&=&
\sum_{m=0}^{\infty}(m+b_{\al})t_m^{\alpha}{\partial\over
\partial t_m^{\alpha}}
+\sum_{m=1}^{\infty}m ({\cal C})_{\alpha}^{\hskip2mm \bt}
t_m^{\alpha}{\partial
\over \partial t_{m-1}^{\beta}}
+{1\over 2\lambda^2}({\cal C})_{\alpha\beta}t^{\alpha}t^{\beta}\nonumber\\
&&
+{1\over 24}\left({3-\dim M\over
2}\chi(M)-\int_Mc_1(M)c_{\dim M-1}(M)
\right)
\ea
where $\chi(M)$ denotes the Euler characteristic of $M$ and the ${\cal C}$ 
is the
matrix of the first Chern class eq.(24). In the case of 
$CP^N$, $({\cal C})_{\al\bt}
=\delta_{\al+\bt+1,N}$  and (29) reduces to (4).

We also recall the recursion relation for descendant two-point functions 
obtained in our previous work \cite{EHX}
\be
(b_{\al}+b_{\bt}+n)\la\sg_n(\Op_{\al})\Op_{\bt}\ra_0=
n\Big(M_{\bt}^{\hskip2mm \gm}\la\sg_{n-1}(\Op_{\al})\Op_{\gm}\ra_0
-\la\sg_{n-1}(c_1(M)\wedge\Op_{\al})\Op_{\bt}\ra_0\Big)
\ee
where
\be
M_{\al\bt}=(b_{\al}+b_{\bt})\la\Op_{\al}\Op_{\bt}\ra_0+({\cal C})_{\al\bt}.
\ee

Let us now derive $L_1$. We first take the $t^{\gm}$-derivative of the 
equation $L_0Z=0$ and keep the genus-0 terms
\ba
&&\sum_{\al=0}^N\sum_{m=0}^{\infty}(m+b_{\al})t_m^{\al}
\la\sg_m(\Op_{\al})\Op_{\gm}\ra_0+b_{\gm}\la\Op_{\gm}\ra_0 \no \\
&&+(N+1)\sum_{\al=0}^N\sum_{m=1}^{\infty}mt_m^{\al}\la\sg_{m-1}
(\Op_{\al+1})\Op_{\gm}\ra_0+(N+1)t_{\gm+1}=0.
\ea
Then multiply (32) by $M_{\bt}^{\hskip2mm \gm}$ and use the recursion 
relation eq.(30). We obtain
\ba
&&\d_{\bt}\hskip1mm \left[\hskip1mm \sum_{\al=0}^N\sum_{m=0}^{\infty}
{(m+b_{\al})(m+b_{\al}+b_{\bt}+1) \over m+1}t_m^{\al}
\la\sg_{m+1}(\Op_{\al})\ra_0 \right. \no  \\
&&+(N+1)\sum_{\al=0}^{N}\sum_{m=0}^{\infty}(2m+2b_{\al}+b_{\bt}+1)
t^{\al}_m\la\sg_m(\Op_{\al+1})\ra_0 \no \\
&&+(N+1)^2\sum_{\al=0}^{N}\sum_{m=1}^{\infty}mt^{\al}_m\la\sg_{m-1}
(\Op_{\al+2})\ra_0
+{1 \over 2}\sum_{\al=0}^N
b^{\al}b_{\al}\la\Op^{\al}\ra_0\la\Op_{\al}\ra_0 \no \\
&&\left.+{1 \over 2}(N+1)^2\sum_{\al=0}^Nt^{\al}t_{\al+2}\hskip1mm \right]
-b_{\bt}(2b_{\bt}+1)\la\sg_1(\Op_{\bt})\ra_0 \no \\
&&-2(N+1)b_{\bt}\la\Op_{\bt+1}\ra_0
+\sum_{\al=0}^Nb^{\al}b_{\bt}\la\Op_{\al}\Op_{\bt}\ra_0\la\Op^{\al}\ra_0=0.
\ea
Note that $\bt$ is not summed in (33). We introduce an auxiliary equation
\be
\tilde{L_0}\hskip1mm : \hskip2mm
\sum_{\al=0}^N\sum_{m=0}^{\infty}t^{\al}_m\la\sg_m(\Op_{\al})\ra_0=2F.
\ee
The above equation follows from the dilaton equation
\be
\la\sg_1(P)\sg_{n_1}(\Op_1)\cdots\sg_{n_s}(\Op_s)\ra_g=
(2g-2+s)\la\sg_{n_1}(\Op_1)\cdots\sg_{n_s}(\Op_s)\ra_g.
\ee
In fact by taking derivatives of $\tilde{L_0}$ and putting all 
the variables $t^{\al}_m=0$ except $t^P_1=-1$ we find
\be
-\la\sg_1(P)\sg_{n_1}(\Op_1)\cdots\sg_{n_s}(\Op_s)\ra_0+
s\la\sg_{n_1}(\Op_1)\cdots\sg_{n_s}(\Op_s)\ra_0
=2\la\sg_{n_1}(\Op_1)\cdots\sg_{n_s}(\Op_s)\ra_0.
\ee
(36) agrees with (35) at $g=0$.
We now take the $t^{\gm}$-derivative of $\tilde{L_0}$ and multiply
$M_{\bt}^{\hskip2mm \gm}$. We obtain
\ba
&&\d_{\bt}\hskip1mm \left[\hskip1mm \sum_{\al=0}^N\sum_{m=0}^{\infty}
{(m+b_{\al}+b_{\bt}+1) \over m+1}t_m^{\al}
\la\sg_{m+1}(\Op_{\al})\ra_0 
+(N+1)\sum_{\al=0}^{N}\sum_{m=0}^{\infty}
t^{\al}_m\la\sg_m(\Op_{\al+1})\ra_0 \right] \no \\
&&\hskip-2mm-2(N+1)\la\Op_{\bt+1}\ra_0-(2b_{\bt}+1)\la\sg_1(\Op_{\bt})\ra_0
-\sum_{\al=0}^N(b_{\al}+b_{\bt})\la\Op_{\al}\Op_{\bt}\ra_0\la\Op^{\al}\ra_0=0.
\ea

Next consider the linear combination $\mbox{(33)}-b_{\bt}\times \mbox{(37)}$
\ba
&&\d_{\bt}\hskip1mm \left[\hskip1mm \sum_{\al=0}^N\sum_{m=0}^{\infty}
{(m+b_{\al}-b_{\bt})(m+b_{\al}+b_{\bt}+1) \over m+1}t_m^{\al}
\la\sg_{m+1}(\Op_{\al})\ra_0 \right.\no  \\
&&+(N+1)\sum_{\al=0}^{N-1}\sum_{m=0}^{\infty}(2m+2b_{\al}+1)
t^{\al}_m\la\sg_m(\Op_{\al+1})\ra_0 \no \\
&&+(N+1)^2\sum_{\al=0}^{N-2}\sum_{m=1}^{\infty}mt^{\al}_m\la\sg_{m-1}
(\Op_{\al+2})\ra_0
+{1 \over 2}\sum_{\al=0}^N
b^{\al}b_{\al}\la\Op_{\al}\ra_0\la\Op^{\al}\ra_0 \no \\
&&\left.+{1 \over 2}(N+1)^2\sum_{\al=0}^{N-2}
t^{\al}t_{\al+2}
+{1 \over 2}b_{\bt}(1+b_{\bt})\sum_{\al=0}^N\la\Op_{\al}\ra_0\la\Op^{\al}\ra_0
\hskip1mm \right]=0.
\ea
Thus the equation becomes a total derivative and we can integrate it 
in $t^{\bt}$ 
(we ignore integration constants). Since the choice of $\bt$ is arbitrary 
in eq.(38), we obtain two equations: one of them comes from the
$\bt$-independent terms and the other one from the terms linear in $b_{\bt}$
(terms quadratic in $b_{\bt}$ give the same equation as the latter),
\ba
&&L_1 \hskip1mm: \hskip2mm
\sum_{\al=0}^N\sum_{m=0}^{\infty}
{(m+b_{\al})(m+b_{\al}+1) \over m+1}t_m^{\al}
\la\sg_{m+1}(\Op_{\al})\ra_0 \no \\
&&+(N+1)\sum_{\al=0}^{N-1}\sum_{m=0}^{\infty}(2m+2b_{\al}+1)
t^{\al}_m\la\sg_m(\Op_{\al+1})\ra_0  \\
&&+(N+1)^2\sum_{\al=0}^{N-2}\sum_{m=1}^{\infty}mt^{\al}_m
\la\sg_{m-1}(\Op_{\al+2})\ra_0
+{1 \over 2}\sum_{\al=0}^N b^{\al}b_{\al}\la\Op_{\al}\ra_0\la\Op^{\al}\ra_0 
\no \\
&&{1 \over 2}(N+1)^2\sum_{\al=0}^{N-2} t^{\al}t_{\al+2}=0, \no \\
&&\tilde{L_1} \hskip1mm:\hskip2mm
-\sum_{\al=0}^N\sum_{m=0}^{\infty}{1 \over m+1}t^{\al}_m\la\sg_{m+1}(\Op_{\al}\ra_0
+{1 \over 2}\sum_{\al=0}^N\la\Op_{\al}\ra_0\la\Op^{\al}\ra_0=0.
\ea

Derivation of $L_2$ is similar. We take the derivative of $L_1$ and multiply
the $M$-matrix and then integrate it with the help of equations derived from
$\tilde{L_0},\tilde{L_1}$. We find $L_2$ and an additional equation 
$\tilde{L_2}$
\ba
&&\hskip-1cmL_{2}\hskip1mm :\hskip2mm 
\sum_{\al=0}^{N}\sum_{m=0}^{\infty}{(m+b_{\al})(m+b_{\al}+1)
(m+b_{\al}+2) \over (m+1)(m+2)} 
t_m^{\al}\la\sg_{m+2}(\Op_{\al})\ra_0 \no \\
&&\hskip-1cm+(N+1)\sum_{\al=0}^{N-1}\sum_{m=0}^{\infty}
{(m+b_{\al})(m+b_{\al}+1)
(m+b_{\al}+2)
\over m+1}
(\sum_{\ell=0}^2{1 \over m+b_{\al}+\ell})t_m^{\al}
\la\sg_{m+1}(\Op_{\al+1})\ra_0\no \\
&&\hskip-1cm+(N+1)^2\sum_{\al=0}^{N-2}\sum_{m=0}^{\infty}
(\sum_{\ell=0}^2(m+b_{\al}+\ell))t_m^{\al}
\la\sg_{m}(\Op_{\al+2})\ra_0+(N+1)^3
\sum_{\al=0}^{N-3}\sum_{m=1}^{\infty}mt_m^{\al}
\la\sg_{m-1}(\Op_{\al+3})\ra_0  \no \\
&&\hskip-1cm +\sum_{\al=0}^{N}{b}^{\al}b_{\al}(b_{\al}+1)
\la\sg_1(\Op_\al)\ra_0
\la\Op^{\al}\ra_0+{N+1 \over 2}\sum_{\al=0}^{N-1}{b}^{\al}b_{\al}(b_{\al}+1)
(\sum_{\ell=-1}^1{1 \over b_{\al}+\ell})\la\Op_{\al+1}\ra_0
\la\Op^{\al}\ra_0  \no \\
&&\hskip-1cm+{1 \over 2}(N+1)^3 \sum_{\al=0}^{N-3} t^{\al}t_{\al+3}=0, \\
&&\hskip-1cm\tilde{L_2} \hskip1mm :\hskip2mm  \no \\
&&\hskip-1cm\sum_{\al=0}^{N}\sum_{m=0}^{\infty}
{b_{\al}+m+1 \over (m+1)(m+2)}t_m^{\al}
\la \sigma_{m+2}({\cal O}_{\al})\ra_0
+(N+1)\sum_{\al=0}^{N-1}\sum_{m=0}^{\infty}{1 \over (m+1)}t_m^{\al}
\la \sigma_{m+1}({\cal O}_{\al +1})\ra_0 \no \\
&&\hskip10mm -\sum_{\al =0}^Nb_{\al}\la {\cal O}^{\al}\ra_0 
\la\sigma_1({\cal O}_{\al})\ra_0
-{N+1 \over 2}\sum_{\al =0}^{N-1}
\la {\cal O}^{\al}\ra_0 \la{\cal O}_{\al +1}\ra_0=0.
\ea

Higher equations $L_n, \hskip1mm n\ge 2$ and their associated ones
$\tilde{L_n}$ will be derived in a similar manner. We postpone our
discussions on $\tilde{L_n}$ and concentrate on
$L_n$ equations. The last step is to convert them 
into differential operators and we find exactly the form of Virasoro 
operators given in eq.(5). We postulate the validity of the Virasoro 
conditions at all genera. Note that the quadratic terms of the correlators 
of the form $\la\cdots\ra\la\cdots\ra$ are promoted to second 
derivative terms in the Virasoro operator which relate amplitudes of 
different genera.

It is interesting to see if we can construct the negative branch 
$L_{-n},\hskip1mm n\ge 2$ of Virasoro operators and compute the central
charge of the algebra. It turns out that in the case of $CP^N$ with $N=$even
it is possible
to construct $\{L_{-n}\}$. They are given by
\ba
L_{-n}\hskip-2mm&=&\hskip-2mm\sum_{m=0}^\infty\sum_{\alpha,j}
(N+1)^jA_\alpha^{(j)}(m,n) t_{m+n+j}^\alpha\d_{m,\alpha+j} \no \\
&&+\frac{\lambda^2}{2}\sum_{\alpha,j}\sum_{m=0}^{n+j-1}
(N+1)^jB_\alpha^{(j)}(m,n) t^\alpha_{n-m+j-1}t_{m,\alpha+j}, \hskip5mm n\ge 1
\ea
The coefficients are defined by
\ba
&&A_\alpha^{(j)}(m,n)\equiv(-1)^j\,\,\,
\frac{(m+1)(m+2)\cdots(m+n+j)}{(b_\alpha+m+j+1)(b_\alpha+m+j+2)
\cdots}\no \\
&&\hskip-5mm\frac{}{\cdots(b_\alpha+m+j+n-1)}
\times\hskip-3mm\sum_{1\le \ell_1\le\ell_2\le\cdots\le\ell_j\le n-1}
\bigg(\prod_{i=1}^j \frac{1}{b_\alpha+m+j+\ell_i}\bigg),
\ea
and
\ba
&&B_\alpha^{(j)}(m,n) \equiv  
(-1)^j
\frac{m!(n-m+j-1)!}{(b^\alpha-j)(b^\alpha+1-j)
\cdots(b^\alpha+m-1-j)} \no \\
&\times&\hskip-4mm\frac{}{(b_\alpha+j)(b_\alpha+1+j)
\cdots(b_\alpha+n-m-2+j)}\sum_{0\le \ell_1\le\ell_2\le\cdots\le\ell_j\le n-2}
\bigg(\prod_{i=1}^j \frac{1}{b_\alpha+j-m+\ell_i}\bigg).\no \\
&&
\ea
We can check the entire algebra and find that the central charge is 
given by
\be
c=N+1.
\ee
This suggests that there exists a realization of our 
algebra by means of $N+1$ free scalar fields. In fact it is possible to 
express Virasoro operators in terms of $N+1$ free scalars and the system
resembles that of a logarithmic conformal field theory.
Details will be discussed elsewhere. In the case of $N=$odd some factors in
the denominators of $A_\alpha^{(j)}(m,n),B_\alpha^{(j)}(m,n)$ vanish and the
above expressions become singular. We do not understand the origin of this
disparity between even and odd values of $N$.

Let us now consider the possible form of Virasoro operators for a general
Fano manifold $M$. A natural conjecture is 
\ba
&&L_n=\sum_{m=0}^\infty\sum_{\alpha,\bt}\sum_j
C_\alpha^{(j)}(m,n) ({\cal C}^{\hskip1mm j})_{\al}^{\hskip2mm\bt}
t_m^\alpha\d_{m+n-j,\bt} \hskip10mm n\ge 1 \\
&&+\frac{\lambda^2}{2}\sum_{\alpha,\bt}\sum_j\sum_{m=0}
D_\alpha^{(j)}(m,n)({\cal C}^{\hskip1mm j})_{\al}^{\hskip2mm\bt}
\d^\alpha_m\d_{n-m-j-1,\bt}
+\frac{1}{2\lambda^2}\sum_{\alpha\bt}
({\cal C}^{\hskip1mm n+1})_{\al}^{\hskip2mm\bt}
t^\alpha t_{\bt}, \no \\
&&L_{-n}=\sum_{m=0}^\infty\sum_{\alpha,\bt}\sum_j
A_\alpha^{(j)}(m,n)({\cal C}^{\hskip1mm j})_{\al}^{\hskip2mm\bt} 
t_{m+n+j}^\alpha\d_{m,\bt},  \hskip10mm n\ge 1\\
&&+\frac{\lambda^2}{2}\sum_{\alpha,\bt}\sum_j\sum_{m=0}
B_\alpha^{(j)}(m,n)({\cal C}^{\hskip1mm j})_{\al}^{\hskip2mm\bt}
t^\alpha_{n-m+j-1}t_{m,\bt},  \no \\
&& b_{\al}=q_{\al}-{(\mbox{dim M}-1) \over 2} \no
\ea
where ${\cal C}^{\hskip1mm j}$ is the $j$-th power of the matrix ${\cal C}$.
(47), (48) reduces to (5), (43) in the case of $CP^N$.
The operators (47), (48) (together with $L_0$ (eq.(29)))
form a Virasoro algebra with a central charge
\be
c=\sum_{\al} 1 =\chi(M),
\ee
if the following condition is satisfied
\be
{1 \over 4}\sum_{\al}b^{\al}b_{\al}
={1 \over 24}\Big({3-\mbox{dim M} \over 2}\chi(M)
-\int_M c_1(M)\wedge c_{dim M-1}(M)\Big).
\ee 
(50) follows from $[L_1,L_{-1}]=2L_0$. 
(Note that we are considering the case where there are no odd-dimensional
cohomologies, $dim H^{odd}(M)=0$ and hence $\sum dim H^{even}=\chi(M)$.)

Eq.(50) is a curious formula depending
only on the geometrical data of $M$. It in fact holds in the case of projective
spaces,
\ba
&&LHS={1 \over 4}\sum_{\al=0}^N\Big(\al+{1-N \over 2}\Big)
(N-\al+{1-N \over 2}\Big)=-{(N^2-1)(N+3) \over 48}, \no \\
&&RHS={1 \over 24}\Big({3-N \over 2}(N+1)-(N+1)^2{N \over 2}\Big)
=-{(N^2-1)(N+3) \over 48}.
\ea 
However, it is possible to show that (50) also holds in other classes of 
Fano varieties, i.e. Grassmannians, 
rational surfaces (point blow-ups of $CP^2$ and $P^1\times P^1$), 
etc. Thus for these
classes of Fano manifolds our Virasoro conditions may also correctly
determine their quantum cohomology. 

We have tested the operators (47) in the case of Grassmannian manifold 
$Gr(2,4)$ for which genus-0
instanton data exist \cite{JS}. We found that Virasoro conditions in fact
reproduce correct instanton numbers of $Gr(2,4)$. Thus we conjecture that 
our Virasoro
conditions are also valid for the Grassmannian manifolds. 
It is a very interesting 
problem to find exactly the class of manifolds for which our construction 
works.

In the above we have concentrated on the discussions of the Virasoro 
conditions. How about the additional constraints $\tilde{L_n}=0$? 
It is easy to check that 
they are in fact satisfied by the genus-0 correlation functions of the $CP^N$
model. It seems that these are the analogues of the W-constraints in the
theory of two-dimensional gravity coupled to minimal models. 
There are, however, higher genus corrections to these equations which we do not
know how to control at present. We would like to have a better 
understanding of these equations in the near future. 

It is quite encouraging for us that the problem of world-sheet instantons 
seem to possess a simple organizing principle 
and the theory has a structure which is a natural
generalization of the 2-dimensional gravity. 
We note that in the case of general target manifold $M$ the central charge 
of the Virasoro algebra is equal to its Euler number (49)
which is the number of supersymmetric vacua of the non-linear $\sg$-model. 
Thus the theory appears to be a free field theory 
of $\chi(M)$ scalar fields each of which describes the fluctuation around
a supersymmetric vacuum.
The presence of the mass gap in the
system may explain the decoupling of different vacua and the free field 
behavior of the theory (private communication 
by Witten). It will be very interesting to see if the free field or
Virasoro structure persists in the case of Fano varieties which have no mass 
gap.

It will also be quite interesting to construct solutions to the Virasoro 
conditions possibly by some matrix 
integrals. In the simplest case of $CP^1$ we already have a matrix model 
with a logarithmic action which reproduces the quantum cohomologies at all 
genera \cite{EY,EHY}. Similar construction 
for more general Fano varieties will be extremely valuable.

\bigskip

We thank discussions with M.Jinzenji and E.Witten. 
Research of T.E. is supported in part by 
Grant-in-Aid for Scientific Research on Priority Area 213 
"Infinite Analysis", Japan Ministry of Education.
Research of K.H. is supported by NSF grant PHY-951497 and DOE grant 
DE-AC03-76SF00098.


\end{document}